\documentclass[preprint,aps,showpacs]{revtex4}

\usepackage{amsmath,graphicx,epsfig,bm}
\begin{document}

\def\pd#1#2{\frac{\partial #1}{\partial #2}}

\title{\bf Motion of a vortex line near the boundary of a
    semi-infinite uniform condensate}
\author {Peter Mason${}^1$, Natalia G. Berloff${}^1$ and Alexander L.
Fetter${}^2$}
\affiliation {${}^1$Department of Applied Mathematics and Theoretical Physics,
University of Cambridge, Wilberforce Road, Cambridge, CB3 0WA, United
    Kingdom\\
${}^2$Departments of Physics and Applied Physics, Stanford University,
    Stanford, CA 94305-4045, USA
}

\date{\today}

\begin {abstract}
We consider  the  motion of a vortex in an asymptotically
homogeneous condensate bounded by a solid wall where the wave
function of the condensate
vanishes. For a vortex  parallel to the wall, the motion is
essentially equivalent to that generated by an image vortex, but
   the depleted surface layer induces an effective shift in the
position of the image  compared to the case of a vortex pair in an
otherwise  uniform flow.
Specifically, the velocity of the vortex can be approximated by $U
\approx (\hbar/2m)\left(y_0-\sqrt 2 \xi\right)^{-1}$,
where $y_0$ is the distance from the center of the vortex to the wall,
$\xi$ is the healing length of the condensate and $m$ is the mass of
the boson.\end{abstract}
\maketitle

\section*{I. INTRODUCTION}
Vortex dynamics in a Bose-Einstein condensate (BEC) has been studied 
intensively, initially in the context of superfluid
helium and later in dilute trapped BECs. The motion of vortices in both uniform
and   inhomogeneous condensates  has been the subject
of many theoretical works, and extensive reviews of these efforts have
been given in \cite{fetterRev,pismen}.

In this paper we consider the problem of  vortex motion in an asymptotically
homogeneous condensate in the
presence of a solid wall where the wave function of the condensate
vanishes. Recent discussions
(see, for example, \cite{anglin} and references therein) on the
motion of vortices near
the surface of trapped condensates have questioned the relevance of
the method of images
   in  describing this motion. In that case, the nonuniform condensate
density is approximated by a linear function that vanishes at the
Thomas-Fermi surface, and the vortex motion can be considered to
arise principally from the local density gradient.  Here, we consider
a rather different situation, in which the condensate density
approaches its bulk value within a healing length $\xi$, and the
vortex is located in the asymptotically uniform region.  In this
latter case, the local gradient of the condensate density is very
small.  The motion can be interpreted as arising from an
   image, but
   the depleted surface layer induces an effective shift in the
position of the image in comparison with the case of  a uniform
incompressible fluid.

Our geometry is  two dimensional, with the vortex aligned along the
$z$ axis, parallel to the surface of the wall.  The dynamics of the
time-dependent BEC in the presence of  the solid
wall at $y=0$ is described by the Gross-Pitaevskii (GP) equation
\begin{equation}
-2 {\rm i} \psi_{t} = \nabla^2 \psi + (1 - |\psi|^2)\psi,
\label{gp}
\end{equation}
subject to the boundary conditions
\begin{equation}
\psi(x,y=0,t)=0, \qquad 0\le y <\infty, \quad |x|<\infty,
\label{boundary}
\end{equation}
in dimensionless units, such that  the distance is measured in healing
lengths $\xi=\hbar/\sqrt{2mgn_0}$, where $g$ is a two-dimensional coupling
constant with dimension of energy times area, $m$ is the mass of the
boson and  $n_0$ is the bulk number density per unit area. Time is
measured in units of $m\xi^2/\hbar$ and energy in units of
$\hbar^2n_0/m$.  In our units,  the speed of sound $c$ in the bulk
condensate is $c=1/\sqrt{2}$.

In the
absence of vortices, the exact solution of (\ref{gp}) for the
stationary state of the semi-infinite condensate    is
\begin{equation}
f(y)=\tanh(y/\sqrt{2}).
\label{f}
\end{equation}
In classical inviscid fluid dynamics with constant mass density
$\rho$,  the relevant kinematic boundary
condition at  a solid wall with  normal vector ${\bf n}$ is
\begin{equation}
\rho \,{\bf u}\cdot {\bf n} =0,
\label{rhou}
\end{equation}
where $\bf u$ is the velocity of the fluid.
The corresponding problem of a vortex moving
parallel to the wall  is solved by placing one or more  image
vortices in such a way that  condition (\ref{rhou}) is  identically
satisfied.

For the dynamics described by the GP equation
(\ref{gp}), the density $\rho \propto |\psi|^2$  is no longer
constant, but rather vanishes at the surface of the wall.    Thus
condition (\ref{rhou}) is automatically satisfied, and all components
of
${\bf u}$ can in principle remain  finite on the boundary.
Therefore, it may seem
that image vortices are irrelevant in the case of the GP equation, so that
   the vortex should remain stationary away from the boundary (where
the fluid density is constant apart from exponentially small
corrections). Our
numerical simulations show that this is not true. In fact, the vortex moves
parallel to the boundary, and it moves {\it faster} than a
corresponding pair of
vortices of opposite circulation in a uniform condensate in the absence
of the depletion caused by the boundary. The purpose of our  paper is
to study this motion in detail.

The paper is organized as follows.
In Sec.\ II we find the  family of  disturbances moving
with a constant velocity along the solid wall by numerically solving
the Gross-Pitaevskii (GP) equation in the frame of reference moving
with the disturbance.  In Sec.\ III a time-dependent Lagrangian
variational analysis is
used to find the first two leading terms in the equation of the vortex
motion in the limit of large distance from the wall. In Sec.\ IV an
alternative approach based on the dependence of total energy and
momentum on the vortex position is used to determine the vortex
velocity. In Sec.\ V we summarize our main findings.
\section*{II. Numerical solutions}
\begin{figure}[t]
\centering
\caption{\baselineskip=10pt \footnotesize [Color online] Graphs of
    the velocity of the vortex $U$ versus the vortex position $y_0$  as
calculated via numerical integration of
    (\ref{ugp}) subject to the boundary conditions $\psi \rightarrow 1$
    as $x^2+y^2 \rightarrow \infty$ without the wall (solid line) and
the asymptotics given by $U=(2 y_0)^{-1}$
    (dashed line). }
\bigskip
\epsfig{figure=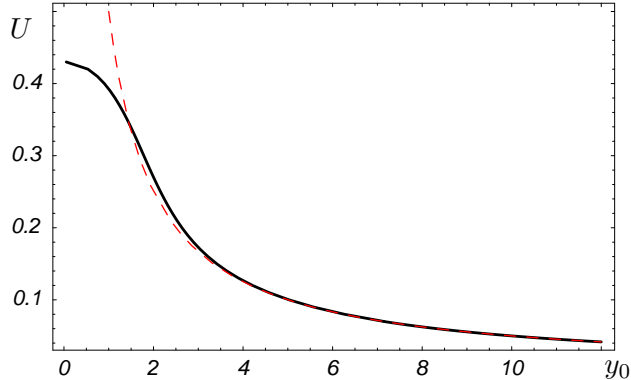,height=2in}
\begin{picture}(0,0)(10,10)
\put(0,13) {$y_0$}
\put(-225,140) {$U$}
\end{picture}
\label{jr}
\end{figure}

In what follows we seek solitary-wave solutions of Eq.\ (\ref{gp}) that
preserve their form as they move parallel to the wall with  fixed
velocity $U$. For each value of the velocity $U$, we have
   $$\psi(x,y,t)=\psi(\eta,y),$$
where $\eta=x-Ut$. The GP equation (\ref{gp}) becomes
\begin{equation}
2 {\rm i} U\psi_{x} = \nabla^2 \psi + (1 - |\psi|^2)\psi,
\label{ugp}
\end{equation}
where we set $x=\eta$.   In the absence of the wall, the
solitary-wave solutions of Eq.\ (\ref{ugp})  were found by Jones and
Roberts \cite{jr4}.  For each value of $U$, there is a well-defined
momentum $p$ and energy $E$, given by
\begin{eqnarray}
p &=&\textstyle{\frac{1}{2{\rm i}}} \int\left[(\psi^*-1)\partial_x\psi
-(\psi-1)\partial_x\psi^*\right]\,dxdy\,,\label{pdef}\\
E &=& {\textstyle{\frac{1}{2}}}\int|\bm\nabla\psi|^2\,dxdy\
+\ {\textstyle{\frac{1}{4}}}\int(1-|\psi|^2)^2dxdy\,.\label{Edef}
\end{eqnarray}
In
a momentum-energy $pE$ plot, the family of  such solitary-wave solutions
   consists of a single branch
that terminates at $p=0$ and $E=0$ as $U\rightarrow c$ (we call this
curve the ``JR dispersion curve'').
For small $U$ and  large $p$ and $E$,
   the solutions are asymptotic to
    pairs of vortices of opposite circulation. As $p$ and $E$ decrease
from infinity, the solutions begin to lose their similarity to
vortex pairs. Eventually, for a velocity $U\approx 0.43$ (momentum
   $p_0\approx 7.7$ and energy $E_0\approx 13.0$)  they lose their
vorticity ($\psi$ loses its zero), and
thereafter the solutions may better be described as
``rarefaction waves'' that can be thought of as  finite amplitude 
sound waves.   The
velocity of the vortex pair in the absence of the boundary is plotted
as
   a function of the position of the vortices $\pm y_0$
   shown in Fig.~\ref{jr}. The dashed line gives the asymptotic velocity
   valid for
    large $y_0$ as $U=(2 y_0)^{-1}$.

\begin{figure}
\centering
\caption{\baselineskip=10pt \footnotesize [Color online] The
    momentum-energy curve of the solitary
    wave solutions of Eq. (\ref{ugp}) with the solid wall (solid line)
    and the JR dispersion curve (dashed line) that has no solid wall.
For the solid-wall boundary condition, the vortex  solutions with
nonzero vorticity and nodes
    are shown in black  and the vorticity-free solutions in
    grey (green).
   }
\bigskip
\epsfig{figure=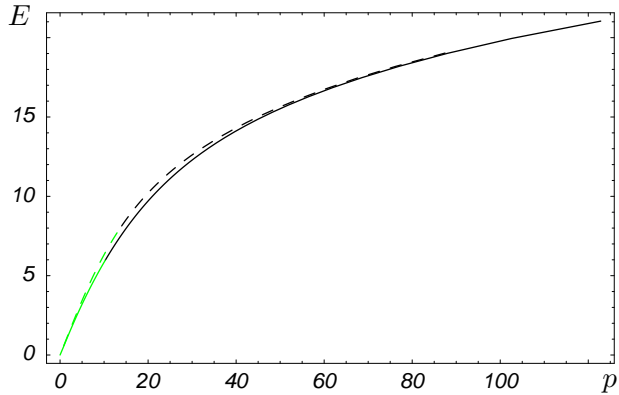,height=2in}
\begin{picture}(0,0)(10,10)
\put(0,13) {$p$}
\put(-225,150) {$E$}
\end{picture}
\label{pemason}
\end{figure}

In analogy with these results, we used numerical methods to find the
complete family of solitary-wave solutions of (\ref{ugp}) subject
to the hard-wall boundary condition (\ref{boundary}). Specifically,
we mapped the semi-infinite domain  onto the box
$(-\frac{\pi}{2},\frac{\pi}{2})\times(0,\frac{\pi}{2})$ using the
transformation
$
\widehat x=\tan^{-1}(D x)
$ and $
\widehat y=\tan^{-1}(D y),$
   where  $D\sim 0.4-1.5$.
The transformed equations  were expressed in a second-order
finite-difference form using $200^2$ grid points, and the resulting
nonlinear
equations were solved by the
Newton-Raphson iteration procedure, using a banded matrix
   linear solver based on the bi-conjugate gradient stabilised
iterative method with
preconditioning. Similar to \cite{jr4}, we are interested in finding
the dispersion curve for our solutions in the $pE$ plane. The energy
and impulse of each
solitary wave  are defined by  the expressions
(\ref{pdef})-(\ref{Edef})
appropriately modified for the ``ground state'' given by $f(y)$:
\begin{eqnarray}
p &=& \textstyle{\frac{1}{2{\rm i}}}\int\left[(\psi^*-f(y))\partial_x\psi
-(\psi-f(y))\partial_x\psi^*\right]\,dxdy\,,\label{pdef2}\\
E &=& \textstyle{{\frac{1}{2}}}\int\left[|\bm \nabla\psi|^2\
+ \textstyle{{\frac{1}{2}}}(1-|\psi|^2)^2-{\rm sech}^4(y/\sqrt{2})\right]\,
dxdy\,.\label{Edef2}
\end{eqnarray}
In Fig.\  \ref{pemason}, we show the resulting solutions in the $pE$
plane.  The plot of the velocity dependence on the
vortex position is given below  in Fig.\ \ref{umason} in Sec.~IV.
All our vortex
solutions with a rigid wall (those with a node in the fluid's
interior) move with
velocities less than  $U=0.47$. For $U>0.47$, the zero of the wave function
occurs on the wall only, and the solitary waves resemble
rarefaction pulses of the JR dispersion curve  away from the wall.

\section*{III. Variational approach}
The time-dependent variational Lagrangian method offers a
convenient analytical approach to estimate  the vortex velocity  for 
large $y_0$.  The  dimensionless GP equation is the
Euler-Lagrange equation for the time-dependent Lagrangian functional
\begin{equation}\label{lagrange}
{\cal L} = {\cal T} - {\cal E}
\equiv \textstyle{\frac{1}{2}} {\rm i} \int  \left(\psi^* \psi_t
- \psi_t^* \psi \right)dxdy -  \textstyle{\frac{1}{2}} \int
\left(|\bm \nabla\psi|^2 + \textstyle{\frac{1}{2}}|\psi|^4\right)dxdy
\end{equation}
where
the time-dependent terms constititute the ``kinetic energy'' $\cal T$
and the remaining terms are the GP energy functional $\cal
E$.

We assume a trial function that depends on one or more parameters, and
use this trial function to evaluate the Lagrangian $\cal L$ in
Eq.~(\ref{lagrange}), which will depend  on the parameters and their
first time derivatives. The resulting  Euler-Lagrange equations
determine the dynamical evolution of the parameters.
  For the present
problem of a vortex moving parallel to a rigid boundary with the
boundary condition (\ref{boundary}), the vortex coordinates
($x_0,y_0$) serve as the appropriate parameters, where
$-\infty<x_0<\infty$ and $0<y_0<\infty$.

When the condensate
contains  a vortex at a distance $y_0$ from the boundary, the
original condensate wave function (\ref{f}) acquires both a phase
$S({\bf r, \bf r}_0)$ and a modulation near the center of the vortex,
where the density vanishes.   To  model this behavior for the half
space, it is preferable to include an image vortex at the image
position $( x_0, -y_0)$.  In this case, the approximate variational
contribution to the phase is
\begin{equation}\label{S}
S({\bf r},
{\bf r}_0) = \arctan\left(\frac{y-y_0}{x-x_0}\right) -
\arctan\left(\frac{y+y_0}{x-x_0}\right),
\end{equation}
where  the
second term reflects the negative image vortex. The derivation of the 
Gross-Pitaevskii
equation involves an integration by parts of the kinetic energy
density $|\bm \nabla \psi|^2 $ to yield $-\psi^*\nabla^2 \psi$ plus a
surface term proportional to $ \psi^* \partial_y\psi$ and this is one 
rationale for
including the image.  Strictly
speaking, this term vanishes because $f(0) $ does so, but the image
vortex ensures that the contribution vanishes even in the case of a
uniform condensate.  The image vortex also cuts off the long-range
tail of the velocity, giving a convergent kinetic energy even for a
semi-infinite condensate. It thus  seems more physical to include the
image in this particular geometry, even though the image is often
omitted for the highly nonuniform density obtained in the
Thomas-Fermi limit for a trapped
condensate~\cite{anglin,al,Kim04,Kim05}.

In addition, the vortex
affects the density near its core, which is modeled by a   factor
$v(|{\bf r} - {\bf r}_0|)$, where $v(r)$ vanishes linearly for small
$r=\sqrt{x^2 + y^2}$ and $v(r) \to 1$ for $r \gg1$~\cite{fe68,al}. 
In principle,  the
function $v(r)$ can be taken as the exact radial solution of the
Gross-Pitaevskii equation in an unbounded condensate, but this choice
requires numerical analysis, and it is often preferable to use a
variational approximation.  A particularly simple choice
is~\cite{Fisc03}
\begin{equation}\label{core2}
v(r) = \begin{cases}
r/\lambda & \text{for $r\le \lambda$};\\

	1  &\text{for $r \ge \lambda$},

	\end{cases}
\end{equation}
where $\lambda$ is the effective
vortex core size;
a variational analysis yields the optimal value
$\lambda = \sqrt 6$.  With these various approximations, the
variational trial function is~\cite{al}
\begin{equation}
\psi({\bf
r}, {\bf r}_0,t) = \,e^{iS({\bf r},{\bf r}_0)}\,f(y)\,v(|({\bf
r}-{\bf r}_0|).
\end{equation}

The time-dependent part of the
functional in Eq.~(\ref{lagrange})  becomes
\begin{eqnarray}
  {\cal
T} & = &-\displaystyle{\int  \left[f(y)\right]^2\partial_tS\,
\left|v(|{\bf r} - {\bf r}_0|)\right|^2}\, dxdy \approx -  \int
\left[f(y)\right]^2\partial_tS \, dxdy,
  \end{eqnarray}
where the last
approximation
omits the effect of the vortex on the density,
replacing  $|v|^2$ by 1 throughout the condensate.   A
straightforward analysis then yields
\begin{equation}\label{calT}
{\cal T} \approx  2\pi \,\dot{x}_0
\int_0^{y_0}   \left[f(y) \right]^2\, dy,
\end{equation}
where
$\dot{x}_0$ is the velocity $U$ of the vortex parallel to the wall.

The  contribution  to $\cal T$ from the vortex core  yields a  term
that is smaller than Eq.~(\ref{calT})  by a factor of relative order
$y_0^{-2}$,  which is negligible relative to  the leading correction
of order $y_0^{-1}$ that we retain here.

Since the energy
functional will  turn out to depend only on the single coordinate
$y_0$,  the Euler-Lagrange equation for $x_0$ implies that $y_0$
remains constant (as expected from energy considerations).  In
contrast, the equation for $y_0$ reduces to
\begin{equation}
\frac{d}{dt}\,\frac{\partial {\cal L}}{\partial
\dot{y}_0} = \frac{\partial {\cal L}}{\partial y_0} = 0,
\end{equation}
since $\dot{y}_0$ does  not appear in ${\cal T}$ (and
hence in $\cal L$).
  Thus the dynamical  motion of the vortex is
given by
   \begin{equation}\label{dynamics}
\dot{x}_0 \approx
\frac{1}{2\pi\left[ f(y_0)\right]^2}  \,\frac{\partial {\cal
E}}{\partial y_0}.
\end{equation}

It is evident that only the
derivative $\partial {\cal E}/\partial y_0$ is relevant, so that
several terms in $\cal E$ play no role in the present analysis.  For
example,  the derivative of the interaction energy ${\cal E}_{\rm
int}(y_0)  = \frac{1}{4} \int |\psi|^4\, dxdy$  vanishes exponentially
for $y_0\gg 1$ and does  not affect the dynamics of the vortex for
large $y_0 $.  Similarly,
the kinetic energy in Eq.~(\ref{lagrange})
separates into two parts, arising from the density variation and  the
flow energy, respectively;  the contribution from the density
variation is also negligible for $y_0\gg 1$.

  The remaining
(dominant)  kinetic energy ${\cal E}_{kv} = \frac{1}{2}\int
|\bm \nabla S|^2\,|\psi|^2\, dxdy$, arises from  the vortex flow.
The
squared velocity now follows from Eq.~(\ref{S})
\begin{equation}
|\bm \nabla S|^2 = \frac{y_0}{y}\left[
\frac{1}{(x-x_0)^2 + (y-y_0)^2} -
\frac{1}{(x-x_0)^2+(y+y_0)^2}\right].
\end{equation}
The resulting
flow-induced kinetic energy is
\begin{equation}
{\cal E}_{kv} =
\frac{1}{2} \int f(y)^2 \,v(|{\bf r-\bf r}_0|)^2 \,|\bm
\nabla S|^2\, dxdy.
\end{equation}
It is convenient to divide  this
integral up into three (strip-shaped) regions
\begin{eqnarray}
\text{region I}: & 0\le y\le y_0-\lambda & \text{
(note that $v = 1 $ in I)}\\
\text{region II}: & y_0 -\lambda \le y\le
y_0+\lambda& \\
\text{region III}: & y_0+\lambda \le y\le \infty  &
\text{(note that $v = 1 $ in III)}
\end{eqnarray}
In region II,
inside the vortex core $|{\bf r - \bf r}_0|\le \lambda$, the
integrals can be found approximately in cylindrical coordinates and
yield ${\cal E}_{kvc}\approx (\pi/2)f(y_0)^2,$ neglecting terms of
order $\lambda ^2/y_0^2$. The remaining region of the strip II
outside the core simplifies because $v = 1$.  It is convenient to
parametrize $y = y_0 +\lambda \sin\theta$ by an angle $\theta$ that
runs from $-\pi/2$ to $\pi/2$.   In this region, the symmetry in $x$
allows us to consider only $x\ge 0$, and the lower limit for $x$ is
$x_m(\theta) = \lambda \cos\theta$.  The relevant integral is
\begin{eqnarray}
&   &{ \displaystyle\int_{x_m(\theta)}^\infty
\left[ \frac{1}{x^2 + (y-y_0)^2} -
\frac{1}{x^2+(y+y_0)^2}\right]}\, dx \nonumber \\
&\qquad=&
{\displaystyle\frac{1}{|y-y_0|}\left[\frac{\pi}{2}-\arctan\left(\frac{x_m(\theta)}{|y-y_0 
|}\right)\right] -
\frac{1}{y+y_0}\left[\frac{\pi}{2}-\arctan\left(\frac{x_m(\theta)}{y+y_0
}\right) \right]}\nonumber \\
&\qquad \approx& {\displaystyle \frac{|\theta|}{\lambda |\sin\theta|}
-\frac{\pi}{4y_0}}.
\end{eqnarray}
The total answer for region II is
\begin{equation}
{\cal E}_{kv\rm II} \approx \pi f(y_0)^2
\left[\frac{1}{2} +\ln 2 -\frac{\lambda }{2y_0} + \cdots\right],
\end{equation}
where $\ln 2$ arises from the definite integral
$\int_{-\pi/2}^{\pi/2} |\theta|/|\tan\theta|\,d\theta = \pi \ln
2$.

In regions I and III, the integrals can be found in cartesian
coordinates, integrating over $x$ first.  Each of these gives two
contributions;  one is simply a combination of logarithms obtained by
replacing $f^2$ by $1$ and the other from the remainder with
$-(1-f^2) = -{\rm
sech}^2(y/\sqrt{2})$.
\begin{eqnarray}\label{I}
{\cal E}_{kv\rm I}
&=&\frac{\pi}{2}\,\ln\left(\frac{2y_0-\lambda}{\lambda}\right) -
\frac{\pi}{2}\int_0^{y_0-\lambda} {\rm
sech}^2(y/\sqrt{2})\,\left(\frac{1}{y_0-y}+ \frac{1}{y_0 +
y}\right)\, dy\\
\label{III}{\cal E}_{kv\rm III} &=&\frac{\pi}{2}\,\left[
2\ln\left(\frac{y_0+\lambda}{\lambda}\right) -
\ln\left(\frac{2y_0+\lambda}{\lambda}\right) \right] \nonumber  \\
&
&- \frac{\pi}{2}\int_{y_0+\lambda}^\infty {\rm
sech}^2(y/\sqrt{2})\,\left(\frac{1}{y-y_0}+ \frac{1}{y +
y_0}-\frac{2}{y}\right)\, dy.
\end{eqnarray}
  To evaluate $\partial {\cal
E}_{kv\rm I}/\partial y_0$ and  $\partial {\cal E}_{kv\rm
III}/\partial y_0$,
we first differentiate the expressions in  Eqs.~(\ref{I})
and (\ref{III}), expand the integrands in the
powers of $1/y_0$ through  $O(y_0^{-2})$ and integrate to get
\begin{eqnarray}
\frac{\partial{\cal E}_{kv\rm I}}{\partial y_0}&\approx&
\frac{\pi}{2 y_0-\lambda} +
    \frac{\sqrt{2}\pi}{y_0^2}{\rm
      tanh}\left(\frac{y_0-\lambda}{\sqrt{2}}\right),\label{7a} \\
\frac{\partial{\cal E}_{kv\rm III}}{\partial y_0}&\approx&
\frac{\pi y_0}{(\lambda + y_0)(\lambda +2
    y_0)}.
   \label{8b}
\end{eqnarray}
A combination of these contributions gives  the vortex velocity in
Eq.~(\ref{dynamics}) as
\begin{eqnarray}
   U=\dot x_0&\approx&\frac{1}{2}{\rm
    coth}^2\frac{y_0}{\sqrt{2}}\biggl[\frac{1}{2y_0-\lambda} +
    \frac{y_0}{(y_0+\lambda)(2y_0+\lambda)}\nonumber \\
    &+&
   \sqrt{2}\biggl(\frac{1}{2}-\frac{\lambda}{2
    y_0} + \ln 2\biggr){\rm sech}^2\bigl(\frac{y_0}{\sqrt{2}}\bigr){\rm
    tanh}\bigl(\frac{y_0}{\sqrt{2}}\bigr) \nonumber \\
&+&\frac{\lambda}{ 2 y_0^2}\,{\rm
    tanh}^2\left(\frac{y_0}{\sqrt{2}}\right)+ \frac{\sqrt{2}}{y_0^2}\,{\rm
    tanh}\left(\frac{y_0-\lambda}{\sqrt{2}}\right)\biggr].
\label{u1}
\end{eqnarray}
This further simplifies to
  \begin{equation}
U\approx\frac{1}{2}\biggl(\frac{1}{2y_0-\lambda} +
  \frac{y_0}{(y_0+\lambda)(2y_0+\lambda)}+\frac{\lambda+2\sqrt{2}}{2y_0^2}\biggr),
\label{u2}
\end{equation}
after we approximate the  hyperbolic functions by their large $y_0$
behavior.  When we neglect terms of order $1/y_0^3$,  the
expression (\ref{u2}) finally reduces to
\begin{equation}
U\approx \frac{1}{2 y_0} \biggl(1 + \frac{\sqrt{2}}{y_0}\biggr),
\label{u3}
\end{equation}
independent of $\lambda$.
\section*{IV. Vortex velocity through the Hamiltonian relationship
    between energy and impulse}

In this section we present a different approach to the asymptotics
for the vortex
velocity based on the relationship between energy and momentum of the
vortex pair.
We compare the motion of a pair of vortices of opposite
circulation (JR solutions) that satisfy
\begin{equation}
2 i U_1 \psi_{1x} = \nabla^2 \psi_1 + (1 - |\psi_1|^2)\psi_2, \qquad
|\psi_1|\rightarrow 1 \qquad
|{\bf x}|\rightarrow \infty,
\label{gp2}
\end{equation}
with the motion of a vortex next to the solid wall
\begin{equation}
2 i U_2 \psi_{2x} = \nabla^2 \psi_2 + (1 - |\psi_2|^2)\psi_2, \qquad
|\psi_2|\rightarrow |\tanh(y/\sqrt{2})| \qquad
|{\bf x}|\rightarrow \infty.
\label{gp3}
\end{equation}
For the asymptotics, we are interested in the solutions for small
$U_i$, for $i=1,2$,
that correspond to a pair of  vortices of opposite circulation. We
   calculate the following quantities: the position of the pair
$(0,\pm y_0)$, the energy and impulse given by
(\ref{Edef})-(\ref{pdef}) for $i=1$ and by
   (\ref{Edef2})-(\ref{pdef2}) for $i=2$,
so that
\begin{equation}
U_i=\frac{\partial E_i}{\partial p_i}.
\label{U}
\end{equation}
These expressions for $i=1$ were derived in \cite{jr4,jr5}; similar
arguments immediately lead to the expressions for $i=2$.

Note that $U_i, E_i $ and $p_i$ are
functions of $y_0$ and if $y_0 \gg 1$,
\begin{equation}
E_1=2 \pi \log(2 y_0), \qquad  p_1=4 \pi y_0,
\label{ep}
\end{equation}
(see, for
instance, \cite{pitaevskii}).
   From (\ref{U}) we have
\begin{equation}
U_1=\frac{\partial E_1/\partial y_0}{\partial p_1/\partial y_0}=\frac{1}{2
    y_0},
\label{uu}
\end{equation}
as expected.

For large $y_0$ an accurate approximation to the solution of
(\ref{gp2}) for the uniform flow was found \cite{b04} as
$\psi_1=u_1(x,y) + i v_1(x,y)$ where
\begin{eqnarray}
u_1(x,y)&=&(x^2+y^2-y_0^2)\tilde R(\sqrt{x^2+(y-y_0)^2})\tilde
R(\sqrt{x^2+(y+y_0)^2}),\nonumber\\
v_1(x,y)&=&-2 x y_0 \tilde R(\sqrt{x^2+(y-y_0)^2})\tilde
R(\sqrt{x^2+(y+y_0)^2}),
\label{uv0}
\end{eqnarray}
where $\tilde R(r)=\sqrt{ (0.3437
+ 0.0286  r^2)/(1 + 0.3333 r^2 +  0.0286  r^4)}.$ Another accurate
choice is
$\tilde R(r)=(r^2 + 2)^{-1/2}$.
Similarly, we expect that $\psi_2$ is accurately approximated by
$\psi_2 =\psi_1
|\tanh(y/\sqrt{2})|$.

\begin{figure}[t!]
\centering
\caption{\baselineskip=10pt \footnotesize [Color online] Graphs of
    the vortex velocity  $U$ versus its distance from the wall $y_0$ as
calculated via numerical integration of
    (\ref{gp3}) (black solid line)  and the asymptotics given by (\ref{u3})
    (red short-dashed line) and by (\ref{ufinal}) (green solid line). Also
    shown is the velocity of the vortex calculated by numerically
    integrating the right-hand side of (\ref{eq1})  (blue long-dashed line).
    As one can see the simplifications made to derive (\ref{ufinal}) are
    consistent with  the full expression (\ref{eq1}) for $y_0>4$. }
\bigskip
\epsfig{figure=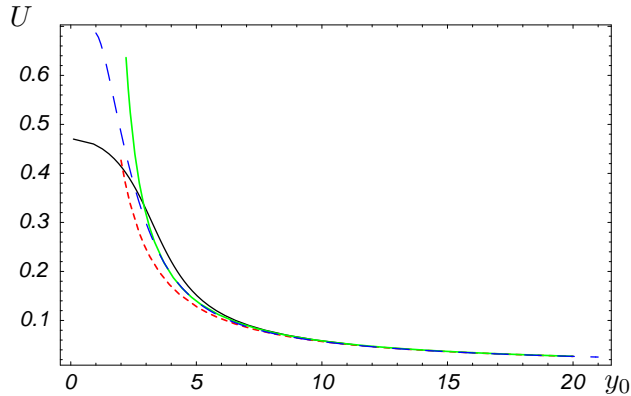,height=2in}
\begin{picture}(0,0)(10,10)
\put(0,13) {$y_0$}
\put(-225,150) {$U$}
\end{picture}
\label{umason}
\end{figure}

The question we pose is:  {\it What is the position of the vortex
    $(0,y_0)$ moving parallel to the solid wall
   with the same velocity as the
vortex pair at $(0,y_0-l)$ in the uniform flow?} Thus we seek the solution of
\begin{equation}
U_1(y_0-l(y_0)) = U_2(y_0) = \frac{\partial E_2/\partial y_0}{\partial
    p_2/\partial y_0},
\label{eq1}
\end{equation}
where we explicitly indicate the dependence of $U_1$ and $U_2$ on the
vortex position. Since $U_1(y_0-l)=(2(y_0-l))^{-1}$, we obtain the
expression for the shift in the vortex position, $l$, in the presence of the
wall as
\begin{equation}
l(y_0)=y_0 - \frac{1}{2}\frac{\partial p_2/\partial y_0}{\partial
    E_2/\partial y_0}.
\label{eq2}
\end{equation}
We rearrange the right-hand side of (\ref{eq2}) and use (\ref{ep}) to
obtain the final
equation that determines $l(y_0)$:
\begin{equation}
l(y_0)=y_0 - \frac{1}{2}\frac{4\pi +\widetilde{dp}}{2 \pi/y_0 +
    \widetilde{dE}},
\label{main}
\end{equation}
where $\widetilde{dE} = \partial(E_2-E_1)/\partial y_0$ and $\widetilde{dp} =
\partial(p_2-p_1)/\partial y_0$ in the integral form given by
(\ref{Edef})-(\ref{pdef}) for $E_1$ and $p_1$ and
(\ref{Edef2})-(\ref{pdef2}) for $E_2$ and $p_2$. In
evaluating the contribution $E_2-E_1$ only the kinetic terms involving
derivatives with respect to $x$ were kept.
   The integrals $\widetilde{dE}$
and $\widetilde{dp}$ are exactly integrable in $x$ with the use of {\it
    Mathematica}, in which   the leading order terms in $1/y_0$ are given by
\begin{eqnarray}
\widetilde{dp} &=& -\frac{8 \pi}{y_0^3}\int_{-\infty}^{\infty}
{\rm sech}^2(y/\sqrt{2})\, dy +
O(y_0^{-5}),\nonumber \\
\widetilde {dE} &=&\biggl(\frac{ \pi}{y_0^2}+\frac{\pi
    (\pi^2-18)}{2y_0^4}\biggr)\int_{-\infty}^{\infty} {\rm
sech}^2(y/\sqrt{2})\, dy +
O(y_0^{-6}).
\label{sol}
\end{eqnarray}
With $\int_{-\infty}^{\infty} {\rm sech}^2(y/\sqrt{2})\,
dy=2\sqrt{2}$ we finally
arrive at
\begin{equation}
l(y_0)= \frac{\sqrt{2} y_0 (\pi^2 + 2(y_0^2-5))}{\sqrt{2} \pi^2 + 2
    (y_0^3 + \sqrt{2}y_0^2 - 9 \sqrt{2})} = \sqrt{2}+O(y_0^{-1}).
\label{ll}
\end{equation}
The vortex next to the wall  moves with the velocity
\begin{equation}
U_2 =\frac{1}{2(y_0-\sqrt{2})},
\label{ufinal}
\end{equation}
which is the main result of our asymptotics. Note that if we expand
(\ref{ufinal}) in a Taylor series we get $U_2=\left (2 
y_0\right)^{-1}\left( 1 +
\sqrt{2}/ y_0\right)$, which agrees with
    the result of Section III.  Fig.~\ref{umason} gives the plot of the
    vortex velocity $U$ as a function of the distance of the vortex from the
    wall $y_0$ for the numerical solutions found in Sec.~II,
    asymptotics found in Sec.~III [see Eq.(\ref{u3})] and asymptotics
    (\ref{ufinal}).

\section*{V.  Discussion and Conclusions}

In a uniform superfluid
with a solid boundary, the  motion of a quantized vortex arises
from
the image that enforces the condition of zero normal flow at the
wall.  In a
trapped condensate, however, the image is generally
omitted.  Instead, the motion  can be
considered to arise from the
gradient of the trap potential, which is the
same as the gradient in
the density in the Thomas-Fermi limit \cite{fetterRev}.  This
behavior
is especially clear for a single vortex at a distance $r_0$
from the center of  a cylindrical container of radius $R$
\cite{Kim04}.  For a classical  incompressible fluid, the vortex
precesses at  an angular velocity
\begin{equation}\label{cl}
\left.
\dot\phi\right|_{\rm cl}  = \frac{\hbar}{m}\,\frac{1}{R^2 -
r_0^2}
\end{equation}
because of the image vortex at $R^2/r_0$.  In
contrast, the precession rate in a trapped cylindrical condensate in
the Thomas-Fermi limit

\begin{equation}\label{TF}
\left.\dot\phi\right|_{\rm TF}  \approx
\frac{\hbar}{m}\,\frac{\ln(R/\xi)}{R^2 - r_0^2}
\end{equation}
is
larger than (\ref{cl}) because of  the (typically large) logarithmic
factor.  Although the denominators of  (\ref{cl}) and (\ref{TF}) both
vary quadratically with $r_0$, the first result arises from the image
and the second from  the parabolic trap potential (and thus the
parabolic density).
If an image were included in the analysis of  the
trap, it would add a correction  of order 1 to the large logarithm
$\ln(R/\xi)$; such a term  is comparable to other terms that are
usually omitted.

As an intermediate situation between these two extremes, the present
paper has  analysed  the dynamics of  a vortex in a half space
bounded by  a
solid wall on which the density of condensate vanishes.  This
geometry represents the simplest problem of a vortex in a condensate
interacting with a surface.   Since the gradient of the density
vanishes exponentially  for $y_0 \gg \xi$,  only the image  remains
to drive the motion in the asymptotic region.
Our  geometry  allows us to
  separate the effect of the
surface from the effect of the density gradient, both of  which appear in the
more complicated problem of  an inhomogeneous trapped condensate
\cite{anglin}.  We
found the complete family of solitary-wave solutions moving with
subcritical velocities parallel to the wall. In addition, both a
variational analysis
and the Hamiltonian relationship between energy and momentum were used
to give the velocity of the vortex as a
function of its distance from the wall.  These results are identical
through  to the first correction  term,
where the small parameter is the inverse distance from the wall. Our
main results
are (i) that the vortex  moves as if there was an image vortex on the
other side of the wall, which essentially replaces the boundary
condition (\ref{rhou}) with  a more stringent requirement  ${\bf
u}\cdot {\bf n} = 0$ and  (ii)
that the depleted surface layer induces an effective shift in the
position of the image in comparison with the case of the uniform flow.
Specifically,   the velocity of the vortex can be approximated by
\begin{equation}
U\approx\frac{ \hbar}{2m(y_0 - \sqrt{2}\xi)},
\end{equation}
where $y_0$ is the distance from the center of the vortex to the wall,
$\xi$ is the healing length of the condensate and $m$ is the mass of
the boson.

\section*{VI. Acknowledgements}

NGB gratefully acknowledges the support from EPSRC. NGB and ALF thank the
organisers of the workshop on Ultracold atoms held at the Aspen Center for
Physics in June 2005, where this work was started, and Eugene Zaremba for a
useful discussion during the workshop.  This work continued at the
Warwick workshop on  Universal features in turbulence: from quantum to
cosmological scales (December 2005); we  thank S. Nazarenko and the
other organizers for their hospitality.

\end{document}